\documentclass[twoside]{dis04}
\def\runauthor{K. Golec-Biernat}
\def\shorttitle{Saturation scale form the BK equation}
\def\be{\begin{equation}}
\def\ee{\end{equation}}
\def\half{\textstyle{1\over 2}}
\def\threehalf{\textstyle{3\over 2}}
\def\alfasbar{\overline{\alpha}_s}
\begin{document}

\title{SATURATION SCALE FROM THE BALITSKY--KOVCHEGOV EQUATION
}

\author{KRZYSZTOF GOLEC-BIERNAT}

\address{H. Niewodnicza\'nski Institute of Nuclear Physics\\
Radzikowskiego 152, 31-342 Krak\'ow, Poland \\
E-mail: golec@ifj.edu.pl }

\maketitle

\abstracts{Recently Munier and Peschanski presented an analysis of the 
Balitsky--Kovchegov (BK) equation concerning the extraction of the saturation scale,
using a simpler equation. We numerically analyze the full BK
equation confirming the universality of their analysis.
}

\section{Introduction}
The Balitsky--Kovchegov equation \cite{1,2} describes saturation \cite{3}
in gamma -- nucleus
deep inelastic scattering, when the structure function 
$F_2\sim \log(1/x_B)$ for the Bjorken variable
$x_B\to 0$.
To be more precise, this behaviour appears in the region where the photon virtuality
$Q^2< Q_s^2(x_B)$. Here $Q_s^2$ is the {\it saturation scale} which rises with
decreasing $x_B$.
Thus the famous BFKL growth \cite{4} in the region $Q^2> Q_s^2(x_B)$, where
$F_2\sim x_B^{-\lambda}$, is tamed due to saturation effects generated  by the BK equation.
The emergence of the saturation scale and its energy dependence is the key issue
in analyses of saturation. Recently, a very elegant method 
of the extraction of the saturation scale was developed in \cite{5}, based on
a simplified version of the BK equation. According to mathematical
properties of this equation,
the obtained results are universal and concern also the full BK equation.
In this presentation we show numerical results which support this claim.

\section{Munier--Peschanski analysis}

In the space of transverse gluon momentum $k$, the BK equation reads
\be
\label{eq:1}
\partial_Y \phi=\alfasbar\chi(-\partial_L)\phi-\alfasbar\phi^2\,,
\ee
where $\phi(Y,L)$
is the Fourier transform of the quark dipole--nucleon scattering amplitude
\cite{2}. Here
$\chi(\gamma)=2\psi(1)-\psi(\gamma)-\psi(1-\gamma)$ is
the BFKL  characteristic function \cite{4}, rapidity $Y=\ln(1/x_B)$,  and
$L=\ln (k^2/k_0^2)$.
The structure function $F_2$ at small $x_B$ can be computed from  solutions
of the BK equation .

Eq. (\ref{eq:1}) is nonlinear, with the linear part given by the BFKL function.
The nonlinearity is responsible for saturation, 
taming the BFKL growth generated by the linear
part alone.
In \cite{5} a remarkable simplification was done approximating the linear part
 by the Taylor expansion
$
\chi(\gamma)\simeq\chi_c+\chi_c^\prime (\gamma-\gamma_c)
+\half\chi_c^{\prime\prime}(\gamma-\gamma_c)^2
$,
 where
$\gamma_c\simeq 0.63$ obeys the equation $\gamma_c\chi_c^\prime=\chi_c$.
Then the linear change of  variables  from
$(Y,L)$ to the new ones $(t,x)$ was performed
\be
\label{eq:2}
t=\half\,(\alfasbar\gamma_c^2\,\chi_c^{\prime\prime})\,Y
~~~~~~~~~~~~~
x=\gamma_c\,L+\alfasbar(\gamma_c^2\,\chi_c^{\prime\prime}-\chi_c)\,Y,
\ee
together with the redefinition:
$u(t,x)={2}/(\gamma_c^2\chi_c^{\prime\prime})\,\phi(Y,L)$. As a result,
the Fisher-Kolmogorov-Petrovsky-Piscunov (FKPP) equation was found,
\be
\label{eq:3}
\partial_t u=\partial_x^2\,u+u(1-u).
\ee
Eq. (\ref{eq:3}) admits {\it traveling wave} solutions:
\be
\label{eq:4}
u(t\to \infty,x)\,\,\sim\,\, u(x-m_\beta(t)),
\ee
where the function $m_\beta(t)$ depends on the asymptotic behaviour
of the initial condition
$u(t_0,x\to \infty)\sim \exp\{-\beta x\}$. For example, for $\beta> 1$
\be
\label{eq:5}
m_\beta(t)=2t-\threehalf\ln t+{\cal{O}}(1)\,.
\ee
This case is particularly important since it contains initial conditions
with color transparency, $\phi\sim 1/k^2$ for $k\to \infty$, corresponding
to $\beta=1/\gamma_c$.
From the form (\ref{eq:4}), the  function $\phi$
has the property of {\it geometric scaling} \cite{6},
\be
\label{eq:6}
\phi(Y,k)=\phi(k^2/Q_s^2(Y)),
\ee
with the {\it saturation scale}  given by
\be
\label{eq:7}
\ln Q_s^2(Y) \,=\,\frac{\alfasbar\chi_c}{\gamma_c}\,Y-\frac{3}{2\gamma_c}\ln Y
-\frac{3}{\gamma_c^2}\sqrt{\frac{2\pi}{\alfasbar\chi_c^{\prime\prime}}}
\frac{1}{\sqrt{Y}}
\,+\,{\cal{O}}(1/{Y})\,,
\ee
where the third term was computed in \cite{7}.
The first three terms are the only universal terms
since the shift $Y\to Y+Y_0$, which
amounts to the change of an initial condition, affects the contribution
${\cal{O}}(1/Y)$.

\section{Numerical analysis}
 The purpose of the numerical
analysis is to confirm that
the results (\ref{eq:6}) and (\ref{eq:7})  are also
valid for the full BK equation.
To this end we numerically solved  eq.~(\ref{eq:1})  with the
help  the Chebyshev polynomial  method.

In Fig.~1 we show the solutions $\phi(Y,k)$ as
functions of transverse gluon momentum $k$ for
twenty different values of rapidity  from $Y_0=0$ to $Y=40$.
The two plots show
the transition from the power-like behaviour $\phi\sim (1/k^\alpha)$ for large $k$
(the r.h.s. plot in the log-log scale)
to the logarithmic behaviour $\phi\sim \ln k$  for $k\to 0$ (the same plot
in the log-lin scale on the l.h.s). This is the illustration of the transition
to saturation due to the nonlinear term in eq.~(\ref{eq:1}), when
the power-like growth
of the gluon distribution for decreasing $k$ is eventually tamed and logarithmic
behaviour sets in. The initial condition (the dashed lines)
contains color transparency
and saturates to a constant value for $k\to 0$. The latter behaviour is immediately changed
by the nonlinear evolution of the BK equation to the logarithmic dependence on $k$.
The FKKP equation, however, conserves the saturation value of the initial
condition producing the traveling wave moving to the right, see \cite{4}.
Thus the truncation of the BFKL kernel in the Munier--Peschanski
analysis influences only the saturation region while for large $k$ the two solutions
are identical.

\begin{figure}[t]
\vspace*{-1.0cm}
\begin{center}
\centerline{\epsfxsize=15cm\epsfbox{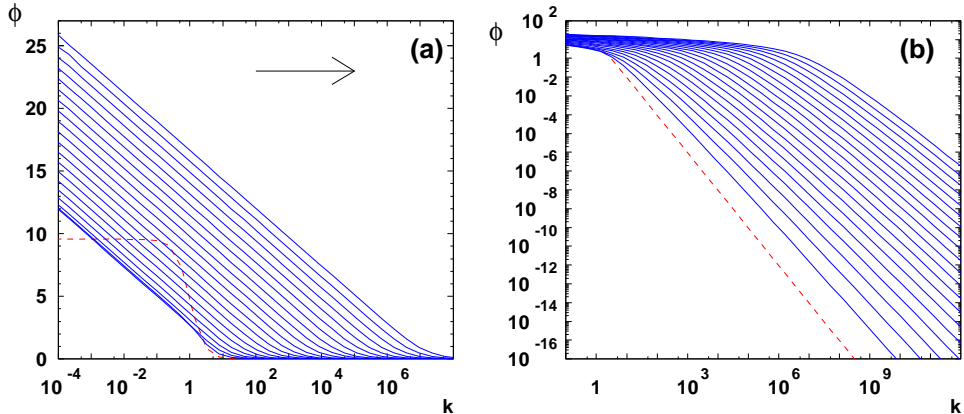}}
\vspace*{-1.0cm}
\caption[*]{The solution $\phi(Y,k)$ of eq.~(1) for twenty values of rapidity
from $Y_0=0$ to $Y=40$ in the order indicated by the arrow. These are the same plots
in the log-lin scale (a), and the log-log scale (b).
They illustrate the transition to saturation with decreasing $k$, when  the
power-like behaviour (b) changes into the logarithmic one (a).
The dashed lines is the initial condition   with color transparency,
$\phi\sim 1/k^2$ for large $k$.
}
\end{center}
\end{figure}

How to extract the saturation scale from numerical solutions ?
The logarithmic
pattern shown in the Fig.~1(a) suggests that after an initial period
of the evolution, when details of the initial condition are washed out, the following
behaviour sets in
\be
\label{eq:8}
\phi(Y,k)=\ln(Q_s(Y)/k)\,.
\ee
This could be checked numerically if the saturation scale computed from (\ref{eq:8}),
$Q_s(Y)=k\exp\{\phi(Y,k)\}$, is independent of $k$. For this purpose we compute
$Q_s(Y)$ for four different values of $\log_{10} k=-2,-1,0,1$. In this range
of $k$ excellent agreement was found, which is shown in Fig.~2(a) where the four
saturation scales coincide.
All momentum scales are in units of some arbitrary scale $k_0$, thus
the absolute normalization of the saturation scale
is not determined. In order to
compare the $Y$-dependence of our saturation scale with the dependence given by
eq.~(\ref{eq:7}),
we compute the derivative $\partial \ln Q_s^2/\partial Y$, which is independent of
the absolute normalization of $Q_s$. The result is shown in Fig.~2(b). The solid lines
correspond to the derivative computed from eq.~(\ref{eq:7}) with one, two or three
first terms in the asymptotic expansion. The lower dashed line is given by differentiation
of the saturation scale defined by eq.~(\ref{eq:8}). We see that such a numerically computed
$Q_s$  agrees with the expansion (\ref{eq:7}) with the
first two terms. The third term is still not visible at the maximal rapidity achieved
in the numerical analysis. It remains to be answered whether this is a numerical effect
or the agreement is achieved at higher values of rapidity.

There is another method of computation of the saturation scale, used in most
of the analyses based on the linear BFKL equation, i.e. from the equation
\be
\label{eq:9}
\phi(Y,Q_s(Y))=\phi_0\sim 1\,.
\ee
This method is more appropriate for the comparison with the Munier--Peschanski
analysis since
it uses the solution in the wave front region. In Fig.~2(b) the upper
dashed curve corresponds to the saturation scale computed
from eq.~(\ref{eq:9}), which agrees very well with
the three term formula (\ref{eq:7}).
\begin{figure}[t]
\vspace*{-1.0cm}
\begin{center}
\centerline{\epsfxsize=15cm\epsfbox{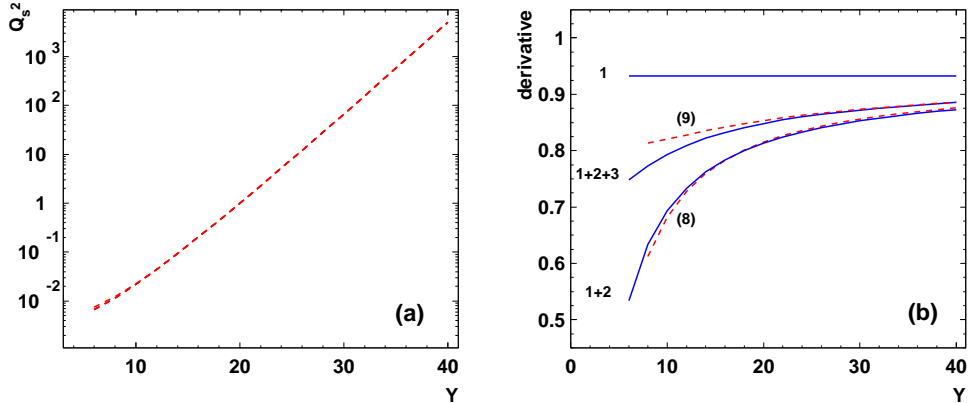}}
\vspace*{-1.0cm}
\caption[*]{In figure (a) the saturation scale computed from (\ref{eq:8}) is shown
for four values of $k$ given in the text with no visible difference between them.
In figure (b) the derivative $\partial\ln Q_s^2/\partial Y$ is shown. The solid
 lines correspond to the derivative of
the analytic formula (\ref{eq:7}) with (as indicated) one, two or three terms.
The lower dashed line is computed from the saturation scale defined by eq.~(\ref{eq:8}),
while the upper dashed line is computed from formula (\ref{eq:9}).
}
\end{center}
\end{figure}

In conclusion,
the analysis of the geometric scaling and saturation scale
in the FKKP equation is general  and applies to the full nonlinear BK equation.

\section*{Acknowledgements}
I thank the organizers for excellent organization of the workshop. Illuminating
discussions with Stephane Munier and Robi Peschanski are gratefully acknowledged.

\end{document}